\documentstyle[12pt]{article}

\newcommand{\F}{\noindent}

\newcommand{\qq}{\qquad}
\newcommand{\q}{\quad}

\newcommand{\MP}{\medskip}
\newcommand{\BP}{\bigskip}

\newcommand{\beq}{\begin{eqnarray}}
\newcommand{\ene}{\end{eqnarray}}
\newcommand{\beqn}{\begin{eqnarray*}}
\newcommand{\enen}{\end{eqnarray*}}

\newcommand{\R}{{\mbox{\bf R}}}
\newcommand{\RR}{{\mbox{\scriptsize {\bf  R}}}}
\newcommand{\C}{{\mbox{\bf C}}}

\newcommand{\HH}{{\cal H}}
\newcommand{\UU}{{\cal U}}

\newcommand{\FF}{{\cal{F}}}

\Large

\begin{document}

\rightline{KIMS-2002-12-23}
\rightline{physics/0212092}
\BP

\vskip12pt

\vskip8pt

\begin{center}
\Large
{\bf Inconsistent Universe

\large
-- Physics as a meta-science --}
\vskip18pt

\normalsize
Hitoshi Kitada
\vskip2pt

Graduate School of Mathematical Sciences

University of Tokyo

Komaba, Meguro-ku, Tokyo 153-8914, Japan

e-mail: kitada@ms.u-tokyo.ac.jp
\vskip10pt
December 23, 2002
\end{center}

\BP

\leftskip24pt
\rightskip24pt

\small

\noindent
{\it Abstract}:
Physics is introduced as a semantics of a formal set theory.

\leftskip0pt
\rightskip0pt

\vskip 24pt

\normalsize

\F
We consider a metatheory of a formal set theory $S$. We name this metatheory $M_S$, indicating that it is a Meta-theory of $S$ as well as a Meta-Scientific theory as Ronald Swan \cite{Swan} refers to. The following arguments are all made in $M_S$.
\BP

We define in $M_S$
$$
\phi =\mbox{ the class of all well-formed formulae of }S.
$$
This $\phi$ is a countable set in the context of $M_S$.
\MP

\F
We identify $\phi$ with the set of truth values (in complex numbers $\C$) of well-formed formulae (wff's) in $\phi$. In this identification, we define a map $T$ from $\phi$ to $\phi$ by
$$
T(\wedge(q))=\mbox{the truth value of a well-formed formula }[\wedge(q)\mbox{ and not }\wedge(q)]
$$
for $q\subset\phi$, with $\wedge(q)$ denoting the conjunction of $q$.
\MP

\F
We note that every subset $q$ of $\phi$ becomes false by adding some well-formed formula $f$ of $\phi$. Hence, the conjunction of $q'=q\cup \{f\}$ is false and satisfies
\MP

$$
T(\wedge(q'))=\wedge(q').
$$

\F
In this sense, $\phi$ is a fixed point of the map $T$.
\MP

\F
Moreover, we have the followings.

\begin{enumerate}
\item
In the sense that any subset $q$ of $\phi$ is false if some well-formed formula is added to $q$, $\phi$ is {\bf inconsistent}.

\item
As $\phi$ is the class of all possible well-formed formulae, $\phi$ is {\bf absolute}.

\item
As $\phi$ is the totality of well-formed formulae, $\phi$ includes the well-formed formula whose meaning is that ``$\phi$ is the class of all well-formed formulae in $S$" in some G\"odel type correspondence between $S$ and $M_S$. In this sense $\phi$ includes (the definition of) $\phi$ itself. Thus $\phi$ is {\bf self-referential} and {\bf self-creative}, and is {\bf self-identical}, just as in M. C. Escher's lithograph in 1948, entitled ``pencil drawing."

\end{enumerate}

\F
The item 3 implies that $\phi$ is a non-well founded set or a hyperset.
\MP

\F
The class $\phi$ is the first world, the Universe, which is completely chaotic. In other words, $\phi$ is ``{\bf absolute inconsistent self-identity}" in the sense of Kitarou Nishida \cite{Nishida}, whose meaning was later clarified by Ronald Swan \cite{Swan} in the form stated above. In this clarification, $\phi$ can be thought ``absolute nothingness" in Hegel's sense.
\MP

\F
The Universe $\phi$ is contradictory, and hence its truth value is constantly oscillating between the two extremal values or poles, truth and false, or $+1$ and $-1$, or more generally, inside a unit sphere of $\C$. Namely, the class $\phi$ as a set of wff's of the set theory $S$ is countable, but the values which the elements of $\phi$ take vary on a unit sphere. In other words, the Universe $\phi$ is a stationary oscillation, when we see its meaning.
\MP

\F
Oscillation is expressed by exponential functions: $\exp(ix\cdot p)$, where $x=(x_1,\cdots,x_d), p=(p_1,\cdots,p_d) \in \R^d$ and $x\cdot p = \sum_{i=1}^d x_i p_i$.
\MP

\F
This $\exp(ix\cdot p)$ is an eigenfunction of the negative Laplacian $-\Delta$:
$$
-\Delta=-\sum_{i=1}^d \frac{\partial^2}{\partial x_i^2}.
$$
Namely
$$
-\Delta \exp(ix\cdot p) = p^2 \exp(ix\cdot p).
$$
\MP

\F
This is generalized to some extent. I.e. if a perturbation $V=V(x)$ satisfies that
$$
H=-\Delta+V(x)\ \mbox{is a self-adjoint operator on}\ \HH=L^2(\R^d),
$$
then
\MP

\F
\begin{center}
$\phi$ is expressed as an eigenfunction of $H$.
\end{center}
\MP

\F
Considering the absolute nature of the Universe $\phi$, we will be led to think that the Hamiltonian $H$ of $\phi$ is a Hamiltonian of infinite degree of freedom on a Hilbert space:
$$
\UU=\{\phi\}=\bigoplus_{n=0}^\infty \left(\bigoplus_{\ell=0}^\infty
\HH^n  \right) \q (\HH^n=\underbrace{\HH\otimes\cdots\otimes
\HH}_{\scriptsize \mbox{$n$ factors}}),
$$
and the Universe $\phi$ is an eigenfunction of the total Hamiltonian $H=H_{\mbox{\scriptsize{\it total}}}$.

Thus we arrive at our first principle.
\BP

\F
{\bf Axiom 1}. The Universe is of infinite nature, and it is eternal.
In other words, the wave function $\phi$ of the Universe satisfies with respect to the total Hamiltonian $H_{\mbox{\scriptsize{\it total}}}$
$$
H_{\mbox{\scriptsize{\it total}}} \ \phi=\lambda\phi
$$
for some non-positive real number $\lambda\le0$.
\BP

In every finite part of $\phi$, a local existence in $\phi$ is expressed by a superposition of exponential functions
$$
\psi(x)=(2\pi)^{-d(N-1)/2}\int_{\RR^{d(N-1)}}\exp(ix\cdot p) g(p)dp
$$
for some natural number $N=n+1\ge 2$ with $n$ corresponding to the superscript $n$ in $\HH^n$ of the definition of $\UU$ above. The function $g(p)$ is called Fourier transform of $\psi(x)$ and satisfies
$$
g(p)=\FF\psi(p):=(2\pi)^{-d(N-1)/2}\int_{\RR^{d(N-1)}}\exp(-ip\cdot y)\psi(y)dy.
$$
A finite subset of wff's in $\phi$ corresponds to a partial Hamiltonian $H$ of $H_{\mbox{\scriptsize{\it total}}}$ of finite degree of freedom, as the content/freedom that is given by a finite number of wff's in $\phi$ corresponds to a finite degree, $n=N-1$, of freedom of a partial wave function $\psi(x)$ of the total wave function $\phi$. If such a partial Hamiltonian $H$ of $H_{\mbox{\scriptsize{\it total}}}$ satisfies some conditions, we can get a similar expansion of a local existence $\psi(x)$ by using generalized eigenfunctions of $H$. This is known as a spectral representation of $H$ in a general setting, but we here are speaking of a more specific expression called generalized Fourier transform or generalized eigenfunction expansion associated with Hamiltonian $H$ (originated by Teruo Ikebe \cite{Ikebe}).
\MP

We call $p$ momentum conjugate to $x$. More precisely we define momentum operator $P=(P_1,\cdots,P_d)$ conjugate to configuration operator $X=(X_1,\cdots,X_d)$ $(X_j=\mbox{multiplication operator by configuration }x_j)$ by
$$
P_j=\FF^{-1}p_j\FF=\frac{1}{i}\frac{\partial}{\partial x_j}\qq(j=1,\cdots,d).
$$
Then $P$ and $X$ satisfy
$$
[P_j,X_\ell]=P_jX_\ell-X_\ell P_j =\delta_{j\ell}\frac{1}{i}.
$$
This shows that what we are dealing with is quantum mechanics. So to be in accordance with actual observation, we modify the definition of $P$
$$
P_j=\frac{\hbar}{i}\frac{\partial}{\partial x_j},
$$
where $\hbar=h/(2\pi)$, and $h$ is Planck constant. Accordingly, the Fourier and inverse Fourier transformations are modified
\beq
&&\FF\psi(p)=g(p)=(2\pi\hbar)^{-d(N-1)/2}\int_{\RR^{d(N-1)}}\exp(-ip\cdot y/\hbar)\psi(y)dy,\nonumber\\
&&\FF^{-1}g(x)=(2\pi\hbar)^{-d(N-1)/2}\int_{\RR^{d(N-1)}}\exp(ix\cdot p/\hbar)g(p)dp.\nonumber
\ene

To sum our arguments up to here, we have constructed quantum mechanics as a semantics of the class $\phi$ of all well-formed formulae of a formal set theory $S$. Quantum mechanics is, in this context, given as an interpretation of set theory.
\BP

We continue to complete our semantics of the Universe $\phi$.
\BP

A local existence is of finite nature, and it is so local that it cannot know the existence of the infinite Universe, and is self-centered. In other words, a local coordinates system starts from its own origin, and it is the self-centered origin of the local system. All things are measured with respect to this local origin.

Therefore we have our second and third principles.
\BP

\F
{\bf Axiom 2}. A local system is of finite nature, having its own origin of position $X$ and momentum $P$, independent of others' origins and others' inside worlds.
\BP

\F
{\bf Axiom 3}. The nature of locality is expressed by a local Hamiltonian
$$
H=-\frac{1}{2}\Delta + V
$$
up to some perturbation $V$, that does not violate the oscillatory nature of local existence. Here $\Delta=\sum_{j=1}^{N-1}\frac{\hbar^2}{\mu_j}\sum_{k=1}^d\frac{\partial^2}{\partial x_{jk}^2}$, the number $N$ corresponds to the number of quantum particles of the local system, and $\mu_j$ is the reduced masses of the particles of the local system.
\BP

A local existence (or local system) is oscillating as a sum or integral of generalized eigenfunctions of $H$. In this sense, the locality or local system is a {\it{stationary oscillating system}}.

A local oscillation may be an eigenfunction of the local Hamiltonian $H$. However, by the very nature that locality is a self-centered existence of finite nature, it is shown that it cannot be an eigenstate of $H$, or more precisely speaking, there is at least one Universe wave function $\phi$ every part of which is not an eigenfunction of the local system Hamiltonian $H$ corresponding to the part. (See \cite{Kitada-localtime}, \cite{Kitada-localsystem}. See also \cite{Kitada-ToL}, \cite{Ki-Fl2}, \cite{Kitada-PI}.)

To express this oscillation explicitly in some ``outer coordinate," we force the locality or local system to oscillate along an ``afterward-introduced" real-valued parameter $t$. The oscillation is then expressed by using the Hamiltonian $H$
$$
\exp(-2\pi itH/h).
$$
This operator is known in QM (quantum mechanics) as the evolution operator of the local system. We call it the local clock of the system, and we call $t$ the local time of the system.

Using our self-centered coordinates of our local system in axiom 2, that is, letting $x$ be position coordinates and $v=m^{-1}P$ be velocity coordinates inside the local system ($m$ being some diagonal mass matrix), we can prove, by virtue of the fact that a local oscillation $\psi(x)$ is not an eigenfunction of $H$, that
$$
\left(\frac{x}{t}-v\right)\exp(-itH/\hbar)\psi(x) \to 0
$$
as $t$ tends to $\pm \infty$ along some sequence in some spectral decomposition of $\exp(-itH/\hbar)\psi$ (see \cite{Kitada-ToL}). This means that the word ``local clock" is appropriate for the operator \linebreak $\exp(-itH/\hbar)$ and so is ``local time" for the parameter $t$. Therefore we also have seen that ``time" exists locally and only locally, exactly by the fact that locality is a self-centered existence of finite nature. This fact corresponds to Ronald Swan's statements in page 27 of \cite{Swan} ``localization must be completely, or unconditionally, circumstantial" and ``localization is not self-creative."

\MP

Once given the local time, the local system obeys Schr\"odinger equation
$$
\left(\frac{\hbar}{i}\frac{d}{dt}+H\right)\exp(-itH/\hbar)\psi(x) =0.
$$

\BP

All up to now can be expressed on a Euclidean space $\R^d$. We need not worry about any curvature as we consider ourselves with respect to our own coordinates.

But when we look at the outside world, our view will be distorted due to the finiteness of our ability. As equivalent existences as localities, we are all subject to one and the same law of distortion.

\MP

Among local systems, we thus pose a law of democracy.

\MP

\F
{\bf Axiom 4}. General Principle of Relativity. Physical worlds or laws are the same for all local observers.
\BP

As a locality, we cannot distinguish between the actual force and the fictitious force, as far as the force is caused by the distortions that our confrontations to the outside world produce.

We have thus the fifth axiom.
\BP

\F
{\bf Axiom 5}. Principle of Equivalence. For any gravitational force, we can choose a coordinate system (as a function of time $t$) where the effect of gravitation vanishes.
\BP

Axioms 4 and 5 are concerned with the distortion of our view when we meet the outside, while axioms 1--3 are about the inside world which is independently conceived as its own. The oscillatory nature of local systems in axiom 3 is a consequence of the locality of the system and the stationary nature of the Universe, so that the oscillation is due to the intrinsic ``internal" cause, while the distortion of our view to the outside is due to observational ``external" cause.

Those two aspects, the internal and the external aspects, are independent mutually, because the internal coordinate system of a local system is a relative one inside the local system and does not have any relation with the external coordinates. Therefore, when we are inside, we are free from the distortion, while when we are meeting the outside, we are in a state that we forget the inside and see a curved world. Thus axioms 1--5 are consistent.

\BP

Quantum mechanics is introduced as a semantic interpretation of a formal set theory, and general relativity is set as a democracy principle among finite, local systems. The origin of local time is in this finitude of local existence, and it gives the general relativistic proper time of each system.

Set theory is a purely inward thought. Physics obtained as semantics of the set theory is a look at it from the outside. The obtained QM itself is a description of the inside world that breeds set theory. The self-reference prevails everywhere and at every stage.


\begin{thebibliography}{88}

\small


\bibitem{Swan} Ronald Swan, {\it A meta-scientific theory of nature and the axiom of pure possibility}, a draft not for publication, 2002.

\bibitem{Nishida} Kitarou Nishida, {\it Absolute inconsistent self-identity} ({\it Zettai-Mujunteki-Jikodouitsu}),
http://www.aozora.gr.jp/cards/000182/files/1755.html, 1989.

\bibitem{Ikebe} T. Ikebe, {\it Eigenfunction expansions associated with the Schr\"odinger operators and their applications to scattering theory}, Arch. Rational Mech. Anal., {\bf 5} (1960), 1-34.

\bibitem{Kitada-ToL} H. Kitada, {\it Theory of local times}, Il Nuovo Cimento
{\bf 109 B, N. 3} (1994), 281-302, http://xxx.lanl.gov/abs/astro-ph/9309051.

\bibitem{Ki-Fl2} H. Kitada and L. Fletcher, {\it Comments on the Problem of Time}, http://xxx.lanl.gov/abs/gr-qc/9708055, 1997.


\bibitem{Kitada-localtime} H. Kitada, {\it A possible solution for the non-existence of time}, http://xxx.lanl.gov/abs/gr-qc/9910081, 1999.

\bibitem{Kitada-PI} H. Kitada and L. Fletcher, {\it Local time and the unification of physics, Part I: Local time},  Apeiron {\bf 3} (1996), 38-45.

\bibitem{Kitada-localsystem} H. Kitada, {\it Local Time and the Unification of Physics Part II. Local System}, http://xxx.lanl.gov/abs/gr-qc/0110066, 2001.

\end{thebibliography}
\end{document}